\magnification=1200
\baselineskip=20pt
\def\dzt{d^z_{\tau}}
\def\dgt{d^{\gamma}_{\tau}}
\def\taum{\tau^-}
\def\taup{\tau^+}
\def\ep{e^+}
\def\em{e^-}
\def\dmugt{\delta\mu^{\gamma}_{\tau}}
\def\dmuzt{\delta\mu^z_{\tau}}
\def\gol{g_{1L}}
\def\gor{g_{1R}}
\def\gtl{g_{2L}}
\def\gtr{g_{2R}}
\def\hol{h_{1L}}
\def\hor{h_{1R}}
\def\htl{h_{2L}}
\def\htr{h_{2R}}
\def\qbcr{{\bar q}^c_R}
\def\tbcl{{\bar t}^c_L}
\def\dbcl{{\bar d}^c_L}
\def\dge{d^{\gamma}_e}
\def\so{S_1}

\def\tilg{{\tilde\gamma}}
\def\mudtt{(\mu_D)_{\tau\tau}}
\def\dat{\delta a_{\tau}}
\def\tmg{\tau\rightarrow\mu\gamma}
\centerline{Dipole moments of tau as a sensitive probe for}
\medskip
\centerline{beyond standard model physics}
\vskip .4truein
\centerline{Uma Mahanta}
\centerline{Mehta Research Institute}
\centerline{10 Kasturba Gandhi Marg}
\centerline{Allahabad-211002, India}
\vskip 1truein
\centerline{Abstract}

CP violating dipole moments of leptons vanish at least to three loop order
and are estimated to be $({m_l\over Mev})\times 1.6\times 10^{- 40}$ e-cm
in the standard model (SM), where $m_l$ is the mass of the lepton.
However they can receive potentially large contributions in some beyond 
SM scenarios and this makes them very sensitive probes of new physics.
In this article we show that a non universal interaction, involving
leptoquarks to the quark-lepton pair of the third generation through
helicity unsuppressed couplings of the order of ordinary gauge couplings, 
can generate electric and weak dipole moments of the order of $10^{-19}$
e-cm for the tau lepton. This is greater than pure supersymmetric (SUSY) and
left-right (L-R) contributions by almost three orders of magnitude.
It is also greater than mirror fermionic contribution by an order of magnitude.
 The measurements
of $\dzt$ and $\dgt$ at LEP, SLC and TCF are expected to reach sensitivities
of $10^{-18}$ e-cm and $10^{-19}$ e-cm respectively in near future. The 
observation of a non vanishing dipole moment of tau at these facilities
would therefore strongly
favour superstring inspired light
leptoquark mediated interactions, over pure SUSY or L-R
interactions and perhaps also mirror generated mixings
without some sort of quark-lepton unification as its origin.
\vfill\eject
\centerline{\bf I.Introduction}

CP violation has so far been observed only in the decay of neutral kaons [1].
In the SM, CP violation arises from the complex Yukawa couplings which
generate a non vanishing phase in the quark mixing 
(Cabibbo -Kobayashi-Maskawa) matrix [2]. On the other hand in leptonic
reactions, CP violation only comes through higher order corrections
involving quark mixing. For the process considered here - the production
of $\tau^+\tau^-$ in $e^+e^-$ collision - the SM prediction for CP
violating effects is so small that it will not be measurable in any 
experiment currently proposed. To give a numerical estimate, CP violating
electric dipole moments (edm's) of leptons vanish at least to three loop
order in the SM and are estimated to be of the order of $1.6({m_l\over Mev})
\times 10^{-40}$ e-cm, where $m_l$ is the mass of the lepton [3].
Observation of CP violation in leptonic systems at current or near future
experimental facilities would therefore signal beyond SM interactions.

From a theoretical standpoint the description of CP violation in the
 framework of the SM does not offer any explanation of its origin. Many
 extensions of the SM have been proposed which try to offer a deeper insight
 into the mechanism of CP violation [4]. Some of them predict CP violating
 effects in interactions where there is no significant contribution from the
 SM. Besides the predicted magnitude of these effects differ from one
extension of the SM to the other, so that their experimental search could
not only detect some beyond SM physics but also shed some light on its nature.
The most sensitive and classic tests in this field are the searches for edm's
of neutron, the electron, the muon and the tau. No non vanishing edm has been
found so far and upper limits have been set at $d^{\gamma}_n<1.1\times 
10^{-25}$ e-cm, $\dge <1.9\times 10^{-26}$ e-cm, 
$d^{\gamma}_{\mu}<1.1\times 10^{-18}$ e-cm and
$\dgt <5\times 10^{-17}$ e-cm [5]. A new search in this field
is the search for a weak diploe (wdm) of $\tau$ ($\dzt$) at the z peak. This
quantity is best constrained from the CERN $e^+e^-$ collider LEP data
to have its real part $(Re \dzt)\le 6.7\times 10^{-18}$ e-cm and an imaginary
part $(Im \dzt)\le 4.5\times 10^{-17}$ e-cm at 95\% CL [6]. These upper limits
were obtained by measuring the expectation values of certain optimal variables
which constitute the dominant CP violating part of the matrix element for
$\tau^+\tau^-$ production. The method of optimal variables is different
from the idea originally proposed by Bernreuther et al which consisted of
measuring CP odd tensor correlations [7] amongst the charged final state
particles in the reaction $e^+e^-\rightarrow \taup+\taum\rightarrow
X^+{\bar\nu}_{\tau}X^-\nu_{\tau}$. One such tensor correlation was looked
for at LEP and 95\% CL limits of Re$\dzt<7.0\times 10^{-17}$ e-cm from
OPAL and Re$\dzt<3.7\times 10^{-17}$ e-cm from ALEPH were obtained from
a sample of 650000 z's [8]. Recently it has been shown [9] that certain CP
odd vector correlations in the reaction
$e^+e^-\rightarrow \tau^+\tau^-\rightarrow X^+{\bar\nu}_{\tau}X^-\nu_{\tau}$
are enhanced significanly when the $e^-$ and $e^+$ beams are longitudinally
polarized. This makes them sensitive to the real and imaginary parts
of the WDFF at the SLAC linear collider (SLC). In the presence of substantial
polarization of $\ep$ and $\em$ beams the same correlations also become
sensitive to the real and imaginary parts of the EDM when the $\taup\taum$
production is no longer dominated by Z exchange, but instead by photon exchange
as in tau-charm factory (TCF). For  a polarization $P_e=\pm .75$ and $10^6$
Z's at SLC these vector correlations could probe $Re \dzt$ and $Im \dzt$
with sensitivities of $3\times 10^{-17}$ e-cm and $1.2\times 10^{-16}$ e-cm
which are comparable with the limits obtained from tensor correlations
and with unpolarized beams.
On the other hand at TCF with 42\% average polarization of each beam and a
total yield of $2\times 10^ 7$ $\taup\taum$ pairs it would be possible to
attain sensitivities of $10^{-19}$ e-cm for $Re \dgt$ and $5\times 10^{-16}$
for $Im \dgt$ with vector correlations. This is an order of magnitude
better than the sensitivity achievable with unpolarized beams.

Theoretically it would be interesting to consider some extension of
the SM  that predicts real parts
of WDFF and EDM at a level that is close to the precision range achievable
at present or in near future experimental facilities. Since $\tau$ belongs
to the third or most massive generation among the fermion families, one
possibility  to generate measurable  dipole moments of $\tau$ would be
to make use of the non-universal interaction that gives rise to large
$m_t$ in loop induced corrections. In this article we shall therefore
consider dipole moments of $\tau$ lepton due to leptoquarks [LQ's] that couple
$\tau$ to t through helicity unsuppressed couplings. Being flavor diagonal
the couplings are not subject to flavor changing suppression either and
can be as large as em coupling [10]. We find that such light ($m_{LQ}\approx
100$ Gev) scalar leptoquarks that couple both to $\tau_L$ and $\tau_R$ with
couplings of magnitude $\vert g_L\vert\approx \vert g_R \vert\approx e$
  can give rise to $\dzt$ and $\dgt$ of the order
 of $10^{-19}$ e-cm for $m_t\approx 175$ Gev. We also find that the pure
 SUSY or L-R contributions to $\dgt$  in models without some sort of quark
 lepton unification are of the order of $10^{-22}$ e-cm. The models considered
 in this article are all invariant under the 
 combined CPT transformation and therefore
 $Im(\dgt)$ and $Im(\dzt)$ turn out to be zero.
  Since the predicted values for $\dgt$ and $\dzt$ due to LQ's
  lie close to the precision range for
measuring these dipole moments at LEP, SLC and TCF, their observation
 in near future would favor such leptoquark scenario over SUSY or L-R
 scenarios without some kind of quark-lepton unification as their origin.
 This in turn would imply some superstring inspired  grand unified
 model like E(6) which can contain such light LQ's without violating
 baryon number and lepton number conservation.
 On the other hand a negative result would not favor any partcular extension
 of the SM. Nevertheless the heirarchy of the predicted values 
 of the dipole moments in different scenarios and the
 proximity of some of them to the current precision range warrants a vigorous
 and continued search for $\tau$ dipole moments to unravel the nature
 of beyond SM physics.
 
 The contents of this article  are divided into the following sections. 
 In Sec. II we present the effective Lagrangian, describing the couplings of
scalar and vector LQ's to quark-lepton pairs, that will be used in this
article to calculate $\dgt$ and $\dzt$. Here we also present the
expressions for CP conserving magnetic moments $(\dmugt, \dmuzt)$ 
and CP violating dipole moments $(\dgt,\dzt)$ due to
$S_1$ type of LQ that incidentally gives rise to the most dominant
 contributions. In Sec. III we present the estimates of the magnitudes and
relative phase of LH and RH leptoquark couplings that will be used in this
 article to calculate
 the dipole and magnetic moments. We also show the consistency of these
 estimates with several pieces of experimental data. In Sec. IV we  estimate
 the dipole moments and magnetic moments of $\tau$ due to LQ's. Here we also
 present the estimates of pure SUSY and L-R contributions to $\dgt$ and compare
 the contributions of different scenarios. In Sec. V we show that for the
parameter values assumed in this article the estimates of $\dgt$ and
$\dmugt$ are consistent with the current experimental limits on
$B(\tau\rightarrow\mu\gamma)$ and $\delta a_{\tau}$ (anomalous magnetic
moment of tau). Finally in Sec. VI we present the conclusions of our study.

\centerline{\bf II. Leptoquark induced dipole moments of tau}

The effective Lagrangian with the most general dimensionless $SU(3)_c\times
SU(2)_l\times U(1)_y $ invariant couplings of scalar and vector
leptoquarks that can give rise to dipole moments of charged leptons can be
written as [10]

$$\eqalignno{L_{eff}&=(\gol\qbcr\imath\tau_2 l_L+\gor\tbcl e_R)S_1
+(\gtl\dbcl\gamma^{\mu}l_L +\gtr\qbcr\gamma_{\mu}e_R)V^+_{2\mu}\cr
&+(\htl{\bar u}_R l_L+\htr{\bar q}_L\imath\tau_2 e_R)R^+_2
+(\hol{\bar q}_L\gamma^{\mu}l_L+\hor{\bar d}_R\gamma^{\mu} e_R)U_{1\mu}\cr
&+h.c.&(1)\cr}$$

Here $q_L, l_L$ are LH quark and lepton doublets, and $e_R, d_R, u_R$ are
RH charged leptons, down and up quarks respectively. $\psi^c$ is a charge
conjugated fermion field. The indices of the LQ's give the dimension of
their $SU(2)$ representation. Color, weak isospin and generation indices
have been suppressed. The subscripts L and R of the coupling constants
stand for lepton chirality.

We shall assume that in the underlying extension of the SM there is some
symmetry that prevents the LQ's from giving rise to baryon and lepton number
violating decays. Such a situation indeed occurs in a four dimensional
E(6) grand unified model derived from a ten dimensional E(8)xE(8)
heterotic superstring theory [11]. There a discrete symmetry arising from
 the topological properties of the compact manifold causes the diquark
 couplings to vanish. 
  The couplings and masses of such LQ's have to
satisfy much weaker bounds. In fact in the low energy superstring models
we obtain relatively small masses for the $\so$ leptoquark ($\approx$ 50-
1000 Gev) [12]. At the CERN large electron-positron collider (LEP) the
experiments have established a lower bound $m_{LQ}\geq 45-73 $ Gev for scalar
leptoquarks [13]. On the other hand, the search for scalar leptoquark
decaying into an electron-jet pair in $p{\bar p}$ colliders have
constrained their masses to be $m_{LQ}\geq 112 $ Gev [14]. Finally the
experiments at the ep collider HERA constrain their masses to be
 $m_{LQ}\geq 92-184$ Gev depending on the leptoquark type and couplings.
  In this article we
shall take the LQ couplings and masses
 to be bounded by low energy processes 
 and by the recent LEP data,
since we do not have a detailed knowledge of the compact manifold where we
realized the compactification. Low energy experiments imply that
  if there is one or more LQ's for each 
quark-lepton generation, it is possible to have flavor diagonal couplings
as large as ordinary gauge couplings for LQ masses of order 100 Gev [10].
Besides the strong helicity suppression on the product $\gol\gor$ or
$\htl\htr$ from the flavor conserving decay $\pi^+\rightarrow e^+\nu_e$
(which implies chiral couplings for LQ's of first generation)
 does not apply for the tau which belongs
to the third generation. Note that LQ's can give rise to CP violating dipole
moments only if they couple to charged leptons of both chiralities.
For scalar LQ's, the dipole moments of $\tau$ get a large contribution from
the chirality flippng top mass in the loop diagram. However for  vector
LQ's, dipole moments of $\tau$ get contribution from the bottom mass in
the loop integral and are therfore much smaller. Besides it is difficult 
to incorporate vector LQ's in a low energy effective theory below 1 Tev.
On the other hand the $SU(2)_w$ singlet, charge 1/3 scalar leptoquark
$\so$ occurs in the superstring inspired E(6) grand unified model. Furher due
to reasons mentioned in Sec. II, they can be relatively light ($m_{\so}
\approx 100$ Gev) without giving rise to proton decay.
 In this article  we shall
therefore consider dipole moments of $\tau$ due to $S_1$ only. The
$SU(2)_w$ doublet, scalar leptoquark $R_2$ gives rise to similar contributions
to $\dgt$ and $\dzt$, but we shall not consider it here. 
 We find that
at one loop order the exchange of $S_1$ leads to the following effective
Lagrangian describing the inetraction of $\gamma$ with the magnetic and
dipole moments of $\tau$

$$L_{eff}=(\imath e/3)N_c m_t I(p,q) \bar{\tau}\sigma_{\mu\nu}[Re(\gol^*
\gor)+\imath \gamma_5 Im(\gol^*\gor)] \tau F^{\mu\nu}.\eqno(2)$$

where q and p are the four momenta of the photon and the incoming $\tau$,
$N_c$ is the number of colors and 

$$I(p,q)=\int (d^4l/(2\pi)^4)
[1/(l^2-m^2_t)((l+q)^2-m^2_t)((l-p)^2-m^2_{S_1})].
\eqno(3)$$

Similarly the effective Lagrangian describing the coupling of Z to the weak
magnetic and dipole moments of $\tau$ turns out to be

$$\eqalignno{L_{eff}&=(\imath e/2 c_w s_w) N_c m_t[({1\over 2}-
{2\over 3}s^2_w)I(p,q)
+{1\over 2}A(p,q)]{\bar\tau}\sigma_{\mu\nu}[Re(\gol^*
\gor)\cr
&+\imath \gamma_5 Im(\gol^*\gor)] \tau Z^{\mu\nu}.&(4)\cr}$$

where $c_w=cos \theta_w, s_w=sin \theta_w $ and

$$A(p,q)q^{\nu}+B(p,q)p^{\nu}=\int (d^4l/(2\pi)^4)
[l^{\nu}/(l^2-m^2_t)((l+q)^2-m^2_t)((l-p)^2-m^2_{S_1})].\eqno(5)$$

I(p,q), A(p,q) and B(p,q) are scalar functions of $p^2, q^2$ and $p.q$.

\centerline{\bf III Estimates of $\vert\gol^*\gor\vert$ and the CP violating
phase $\delta$}

Since the LQ's considered in this article do not lead to baryon or lepton
number violating decays, in general
their couplings of either helicity (but not both)
can be of the order of em coupling if their masses are of the order of
100 Gev. The restriction to couplings of either helicity but not both,
arises from the helicity suppressed deacy $\pi^+\rightarrow e^+\nu_e$
and applies only to LQ couplings to the quark-lepton pair of
 first generation [10].
In fact one finds that $\vert \gol\gor\vert ^{1\over 2}<{m_{S_1}\over 10
Tev}$ for LQ couplings to the fermions of first generation.
 However this helicity
constraint does not apply for the tau  since it is quite massive and
both $\vert\gol\vert$ and $\vert\gor\vert$ can be simultaneously of the
order of e. The recent LEP data on $z\rightarrow \taup\taum$ decay
can be used to impose constraints on the masses and couplings for  the
third generation LQ's which couple to the top quark. It would be interesting
to examine the implications of those constraints on the dipole moments
of $\tau$ lepton. Mizukoshi et al. [16] evaluated the one-loop contribution
due to LQ's to all LEP observables and made a global fit to extract 95\%
confidence level limits on the LQ masses and couplings. The limits
obtained by them are most stringent for LQ's that couple to the top
quark since their contributions are enhanced by powers of the top quark mass.
Moreover, the limits are slightly better for LH couplings than for RH
couplings for a given LQ. From the allowed region in the $m_{LQ}-\gol(\gor)$
plane for $\so$ LQ, we find that for $m_{LQ}\approx 100 $ Gev, the LEP
limits are $\vert\gol\vert<.5$ and $\vert\gor\vert<.5$. The values of LQ
mass ($m_{LQ}\approx 100$ Gev) and couplings
 ($\vert\gol\vert\approx\vert\gor\vert\approx e\approx .3$)
 assumed by us in this article are therefore
close to and consistent with the limits implied by LEP data. 

The CP violating phase $\delta$ (where we define $\gol^*\gor =
\vert\gol^*\gor\vert e^{\imath\delta} $) can be estimated or rather an
upper limit on it can be derived from the experimental limit on $d^{\gamma}_e$.
In order to do that we shall assume that the phase $\delta$ for the third
generation is of the same order as that of the first generation. Any heirarchy
in the dipole moments of leptons of different generations will arise from
the chirality flipping mass in the loop diagram and from the constraint on
flavor changing LQ couplings. Under these assumptions we find that (using
naive dimensional anlysis)

$$d^{\gamma}_e\approx -{2\over 3}e N_c\vert\gol^*\gor\vert\sin\delta{\xi\over
16\pi^2}{m_u\over m^2_{S_1}}.\eqno(6)$$

where $\xi$ is a number of O(1) which arises in evaluating the loop integral.
Note that for the couplings of $\so$ to the quark-lepton pair of first
generation $\vert \gol^*\gor \vert\le {m_{\so}\over 10 Tev}$.
From the experimental limit  $\dge\le 2\times 10^{-26}$ e-cm [5], it
then follows that $\sin\delta\le{5\over 6.3\xi}\approx O(1)$. This implies
that $cos\delta\approx O(1)$ and therefore LQ contribution to $\dmugt$
turns out to be of the same order as $\dgt$. The fact that $\dmugt$
is of the same order as $\dgt$ makes the former
 very small and consistent with the
experimental limit on $\vert a^{exp}_{\tau}-a^{sm}_{\tau}\vert$, where
$a_{\tau}$ is the anomalous magnetic moment of $\tau$.

\centerline{\bf IV. Estimates of dipole moments and magnetic moments of $\tau$}

\centerline{\bf in different scenarios}

In order to evaluate $\dgt$ and $\dmugt$ we have to find the value of
I(p,q) for $p^2=m^2_{\tau}$ and $-2p.q=q^2=0$ corresponding to on shell
$\tau $ and $\gamma $. On the other hand to find $\dzt$ and $\dmuzt$  we
have to find the values of I(p,q) and A(p,q) for $p^2=m^2_{\tau}$ and
$-2p.q=q^2=M^2_z$ corresponding to on shell $\tau$ and Z. From the
expressions for the effective Lagrangians
given in Sec. III we find that

$$\eqalignno{\dmugt &=-(2\imath e/3)N_c m_t I(p,q)Re(\gol^*\gor)\approx
1.1\times 10^{-19} e-cm.&(7a)\cr
\dgt &=-(2\imath e/3)N_c m_t I(p,q)Im(\gol^*\gor)\approx
1.1\times 10^{-19} e-cm.&(7b)\cr}$$
and

$$\eqalignno{\dmuzt &=-(\imath e/c_w s_w)N_c m_t [({1\over 2}-{2\over 3}s^2_w)
I(p,q)+{1\over 2}A(p,q)]Re(\gol^*\gor)\cr
&\approx 2\times 10^{-19}e-cm.&(8a)\cr
\dzt &=-(\imath e/c_w s_w)N_c m_t [({1\over 2}-{2\over 3}s^2_w)
I(p,q)+{1\over 2}A(p,q)]Im(\gol^*\gor)\cr
&\approx 2\times 10^{-19}e-cm.&(8b)\cr}$$

for $m_t\approx 175$Gev, $m_{\so}\approx 100$ Gev, $\vert\gol^*\gor\vert\approx
e^2\approx .1$ and $\cos\delta\approx \sin\delta\approx O(1)$. The
relevant loop integrals appearing in the above expressions have been
evaluated numerically. An order of magnitude estimate of the magnetic moments
and dipole moments can also be obtained by naive dimensional analysis.
 
 We will now consider the electric dipole moments of $\tau$ in SUSY and
 L-R symmetric models which do not incorporate some sort of quark-lepton
 unification and hence do not have LQ's. In SUSY models there can be
inherently supersymmetric contributions that are large. For example at one
loop level a non-vanishing dipole moment can arise from a $\tau$ going
into a scalar $\tau$ and a neutralino
$(\tilg,{\tilde Z},{\tilde H})$. For the photino mediated diagram
 we get ${\dgt\over e}={\alpha\over \pi}{m_{\tau}Im(Am_{\tilg})
 \over m^3} f(x)$ where $m_{\tilg}$ is the mass of the photino; m is the
scale for low energy SUSY breaking; $x=(m_{\tilg}/m)$; A is a complex parameter
of order unity and f(x) is the Polchinski-Wise function [13]. The CP
violating phase $arg(Am_{\tilg})$ arises from the effective soft SUSY
breaking terms and can be bounded from $\dge$ [14]. Using the experimental
bound $\dge<2\times 10^{-26}$ e-cm we find $arg(Am_{\tilg})<10^{-2}$.
Hence $\dgt\le 7.2\times 10^{-23}$e-cm. Inherently supersymmetric
contribution to $\dgt$ is therefore less than the LQ contribution
 by almost three orders of magnitude. 

 For L-R symmetric model a sizeable $\dgt$ can arise if the $\tau$ couples
to a RH heavy neutrino. Chang, Pal and Nieves [15] find that for each 
$W_i$ and $\chi_A$ running in the loop ($W_i$ and $\chi_A$ are mass
eigenstates for charged gauge bosons and neutrinos repectively)

$$d^{iA}_{\tau}=-{eg^2m_A\over 64\pi^2M^2_i}U_{Li}U_{Ri}Im(P_{3A}Q_{3A})
[{r^2_{Ai}-11r_{Ai}+4\over(r_{Ai}-1)^2}+{6r^2_{Ai}\ln r_{Ai}\over
(r_{Ai}-1)^3}].\eqno(9)$$
where $m_A=$ mass of $\chi_A$; $M_i=$ mass of $W_i$ and
 $r_{Ai}={m^2_A\over M^2_i}$.
U and P,Q are unitary matrices that relate gauge eigenstates to the mass
eigenstates of the charged bosons and neutrinos respectively. For $M_2\gg
M_1$ we get

$$\dgt\approx(1.0\times 10^{-24}e-cm)\sin 2\zeta\sum_A ({m_A\over 1Mev})
Im(P_{3A}Q_{3A})D_A.\eqno(10)$$

where $D_A$ is the expression in brackets in Eq. (9) and $\zeta$ 
is the $W_L-W_R$ mixing angle. From current algebra analysis of purely
non-leptonic strange decays one obtains $\zeta<.004$. Hence $\dgt\le
10^{-26}{Im\mudtt\over 1Mev}$, where $\mu_D$ is the Dirac mass in the neutrino
mass matrix. The upper bound on $Im\mudtt$ can be estimated
from the LH neutrino mass $m_{\nu_{\tau}}\approx {\mudtt^2\over \mu_N}$,
where $\mu_N$ is the mass of the RH neutrino.
For $\mu_N\approx 1$ Tev and $m_{\nu_{\tau}}^{exp}\le 35$ Mev we get
$Im\mudtt\le 6$ Gev. Thus $\dgt\le 2.4\times 10^{-22}$ e-cm. Inherently
L-R contribution to $\dgt$ is therefore also less than the LQ contribution
by three orders of magnitude.

The EDM's of ordinary fermions can also receive large contributions if the
theory contains mirror fermions. The EDM's in this case arise from mixings
between ordinary and exotic mirror fermions. Very tight bounds have been
placed on these mixings by Langacker and London [20] from a combined
analysis of various experiments and by Bhatacharya et al. [21] using the
LEP data. Using these constraints on mixings and maximizing the CP violating
phase, one can derive limits on mirror fermionic contributions to various
dipole moments. In particular using the limits on the mixing angles from
 the LEP data on $z\rightarrow\taup\taum $ decay, Joshipura [22] has shown
that a $\dgt$ of the   order of $2.1\times 10^{-20}$ e-cm can be
generated. Note first that the $\dgt$ generated by mirror fermions
is still an order of magnitude smaller than the leptoquark contribution
derived by us in this article. Second, although mirror fermions can occur
in a low energy effective theory which does not incorporate any
 quark-lepton unification, their existence becomes natural in the context
of grand unified models based on large enough orthogonal groups.

\centerline{\bf V. Leptoquark contribution to B($\tau\rightarrow \mu\gamma$) 
and $\dat$ }

The rare decay $\tmg$ takes place through the transition magnetic and electric
dipole moments between $\tau$ and $\mu$. Leptoquark contributions to B($\tmg$)
and $\dgt$ can therefore be related. It can be shown that

$$\dgt=[B(\tmg)]^{1\over 2}(Gm_{\tau}/2{\sqrt 6}\pi e)\delta'.\eqno(11)$$

where for the $\so$ leptoquark with the top quark contribution dominating

$$\delta'\approx {\sqrt 2}Im[(\gol^*)_{33}(\gor)_{33}]/[\vert (\gol^*)_{32}
(\gor)_{33}\vert^2+\vert (\gol^*)_{33}(\gor)_{32}\vert^2]^{1\over 2}.
\eqno(12)$$

Here the subscripts i,j of the couplings denote the generations
to which the quark and lepton belong. For $\vert (\gol^*)_{33}\vert\approx
\vert (\gor)_{33}\vert\approx {1\over 3}$ and
$\vert (\gol^*)_{32}\vert\approx\vert (\gor)_{32}\vert\approx 10^{-3}$
(which is the present bound on flavor changing LQ couplings) [10] we find that
$\delta'\approx ({\sqrt 2}/ 3)10^3 \sin\delta$ where $\delta$ is the CP
violating phase. From the experimental upper limit [5] B$(\tmg)\le 4\times
10^{-6}$ it then follows that $\dgt\le 7.2\times 10^{-19}$ e-cm. Hence the
upper limit on $\dgt$ derived from B$(\tmg)$ and some plausible assumptions
regarding the magnitude of flavor changing couplings is consistent with
the limit obtained from $\dge$. Alternatively for the values of the parameters
assumed in this article our LQ model predicts a value for B($\tmg$) that
is close to the present experimental bound.

In our LQ model the contributions to $\tau$ lepton's anomalous magnetic moments
$\dmugt$ and $\dmuzt$ turn out to be of the same order as the CP 
violating dipole moments
$\dgt$ and $\dzt$. This makes $\dmugt$ and $\dmuzt$ quite small, since the
scale of CP violating dipole moments are naturally small. Nevertheless it
would be interesting to see if the estimate of $\dmugt$ is consistent with
the experimental bound on $\dat =\vert a^{exp}_{\tau}-a^{sm}_{\tau}\vert $.
The contribution of $\so$ leptoquark to $\dat$ is given by  $\dat\approx
2.4 N_c Re(\gol^*\gor)m_{\tau}m_t\times 10^{-7}\approx 1.9\times 10^{-5}$
if we take $\cos\delta\approx O(1)$. The SM model contribution [5] is
given by  $a^{sm}_{\tau}\approx .0011773$. Whereas the experimental limit
[5] is $a^{exp}_{\tau}\approx .01$ at 95\% CL. The LQ conyribution to $\dat$
is therefore well within the experimental bound of $\dat\le.01$

\centerline{\bf VI. Conclusions} 
 
 In this article we have considered a particular species of light 
scalar LQ which occur in superstring inspired E(6) grand unified model
and whose contributions to the electric and weak dipole moments of $\tau$
lepton are of the order of $10^{-19}$ e-cm. This should be compared with
the precision range $\delta Re(\dgt)\le 10^{-19}$ e-cm and 
$\delta Re(\dzt)\le 10^{-18}$ e-cm. that can be achieved at current or propsed
experimental facilities. The closeness of the predicted values to the
precision range makes both this scenario and the experimental search of
dipole moments of $\tau$ interesting and worth pursuing. We have estimated
the dipole moments for LQ couplings whose magnitudes are of the order
 of em coupling and relative phase of $O(1)$. For smaller values of these
 parameters the estimated dipole moments would be smaller. We have also
shown that the inherently SUSY and L-R contributions to the EDM of $\tau$ are
of the order of $10^{-22}$ e-cm, which is too small to be observed in any
of the proposed experiments. The mirror fermionic contribution to
$\dgt $, which is of order $10^{-20}$ e-cm, is also smaller than the
projected precision range of the future experimental facilities by an
order of magnitude.
 The observation of a non-vanishing dipole
moment of $\tau$ in near furure would therefore favor light LQ mediated
interaction over pure SUSY or L-R interactions and perhaps also mirror
generated mixings
 (without some sort of quark lepton
unification) as its origin. This in turn would imply some  superstring
inspired grand unified model like E(6). Dipole moments of $\tau$ could
therefore be used as a sensitive probe for unravelling the nature of beyond
SM physics. Thus a strong search program for dipole moments of $\tau$
at LEP, SLC and TCF is stongly warranted. Finally we have shown that our
estimates of $\dgt$ and $\dmugt$ are consistent with the experimental
constraints on $B(\tmg)$ and $\dat$. An interesting feature of our
estimates is that the CP conserving magnetic moments turn out to be 
of the same order as the CP violating dipole moments if we assume 
$\sin \delta\approx \cos\delta\approx O(1)$. Further since the LQ
interactions considerd in this article are invariant under the CPT
transformation, the imaginary parts of the dipole moments necessarily turn
out to be zero. A CPT odd observable O can have a nonzero expectation value
only in the presence of an absorptive part of the amplitude. Since the final
state interactions, which could give rise to an absorptive part, is
negligible in weak $\tau$ decays, O must be proportional to Im$\dgt$
or Im$\dzt$. Measurement of some CPT odd quantity can therefore be used to
search for imaginary parts of dipole moments of $\tau$ and verify the 
prediction of our LQ model. 

\centerline{\bf References}

\item{1.} J. H. Christension et al., Phys. Rev. Lett. 13, 138 (1964).

\item{2.} M. Kobayashi and T. Maskawa, Prog. Theor. Phys. 49, 652 (1973).

\item{3.} M. J. Booth, University of Chicago Report No. EFI-93-02,
hep-ph/9301293 (unpublished).

\item{4.} "CP Violation", Advanced series on Directions in High Energy
Physics- vol 3, editor C. Jarlskog, World Scientific.

\item{5} Particle Data Group, Phys. Rev. D 50, 1173 (1994).

\item{6.}OPAL collaboration, R. Akers et al., CERN Report No.
 CERN-PPE/94-171 (unpublished).
 
 \item{7.} W. Bernreuther and O. Nachtmann, Phys. Rev. Lett. 63, 2787 (1989);
 W. Bernreuther et al., Z. Phys. C 52, 567 (1991).
 
 \item{8.} OPAL collaboration, P. D. Acton et al., Phys. Lett B 281, 405
  (1992); ALEPH Collaboration, D. Buskulic et al., ibid 297, 459 (1992).
  
  \item{9.} B. Ananthanarayan and S. D. Rindani, Phys. Rev. Lett. 73,
  1215 (1994); Phys. Rev. D 51, 5996 (1995).
  
  \item{10.} W. Buchmuller and D. Wyler, Phys. Lett. B 177, 377 (1986);
  ibid 191, 442 (1987).
  
  \item{11.} E. Witten, Nucl. Phys. B 258, 75 (1985);
   M. Dine , V. Kaplunovsky, M. Mangano, C. Nappi and N. Seiberg, Nucl. Phys.
   B 259 (589) 1985.
 
 \item{12.} J. Ellis,, K. Enqvist, D. V. Nanopoulos and F. Zwirner, CERN
 preprint TH.4323/85 (1985); L. E. Ibanez and J. Mas, CERN preprint 
 TH.4426/86 (1986).
 
 \item{13.} L3 Collaboration, B. Adeva et al., Phys. Lett B 261, 169 (1992).
 
 \item{14.} CDF Collaboration, F. Abe et al., Phys. Rev D 48, 3939 (1993).
 
 \item{15.} ZEUS Collaboration, M. Derrick et al., Phys. Lett B 306,
 173 (1993).
 
 \item{16.} J. K. Mizukoshi, O. J. P. Eboli and M. C. Gonzalez-Garcia,
 Nucl. Phys. B 444, 20 (1995).
           
 \item{17.} J. Polchinski and M. B. Wise, Phys. Lett B 125, 393 (1983).
  
  \item{18.} S. M. Barr and A. Masiero, Phys. Rev. Lett. 58, 187 (1987).
  
  \item{19.} J. F. Nieves, D. Chang and P. B. Pal, Phys. Rev. D 33,
  3324 (1986).
  
  \item{20.} P. Langacker and D. London, Phys. Rev. D 38, 886 (1988);
  Phys. Rev. D 38, 907 (1988).
  
  \item{21.} G. Bhatacharya et al., Phys. Rev. Lett. 64, 2870 (1990).
  G. Bhatacharya et al., Phys. Rev. D 42, 268 (1990).
  
  \item{22.} A. S. Joshipura, Phys. Rev. D 43, 25 (1991).
  
\end